# Modeling Apsidal Motion in Eclipsing Binaries Using ELC

Alexander J. Dimoff[1] and Jerome A. Orosz[2]
[1] Max Planck Institute for Astronomy, Königstuhl 17, D-69117 Heidelberg, DE, Germany; dimoff@mpia.de
[2] Department of Astronomy, San Diego State University, 5500 Campanile Drive, San Diego, CA 92182, USA

## Abstract

Apsidal motion is the precession of the line of apsides in the orbit of a binary star due to perturbations from General Relativity (GR), tides, or third-body interactions. The rate of precession due to tidal effects depends on the interior structures of the stars, and as a result, binaries in which this precession occurs are of great interest. Apsidal motion is observed through the analysis of eclipse times, which reveal small changes in the average interval between successive primary and secondary eclipses, taking all available observed times of eclipse and yielding an estimate of the apsidal rate. Given that this is a single observed quantity, various degeneracies are unavoidably present. Ideally, one would have a model that predicts eclipse times given the orbital and stellar parameters. These parameters for a given binary could then be computed using least squares, provided a suitably large number of eclipse times. Here we use the eclipsing light curve (ELC) program as such a model. The Newtonian equations of motion with additional force terms accounting for GR contributions and tidal distortions are integrated, yielding precise sky positions as a function of time. Times of mid-eclipse and instantaneous orbital elements are computed as a function of time. In this paper, we outline the method and compare numerically computed apsidal rates with standard formulae using a set of 15 binaries based on real systems. For our simulated systems, the derived apsidal rates agree with the standard formula.

*Unified Astronomy Thesaurus concepts:* Apsidal motion (62)

## 1. Introduction

Apsidal motion is the precession of the line of apsides in an orbit, generally in the same direction as the motion of the secondary body. For a binary with an eccentric orbit, the phase difference between the primary and secondary eclipse depends on the argument of periastron $\omega$. If $\omega$ changes over time, then the primary and secondary eclipse times will not follow simple linear ephemerides. When the primary and secondary eclipse times are fit to a common ephemeris (the free parameters to this fit are the common period, the common reference time of primary eclipse, and the average time difference between the primary and secondary eclipses), the residuals in the common period observed minus computed (CPOC, or $O - C$) diagram are roughly sinusoidal with opposite phases (see Figure 1 for a schematic example for a binary resembling Y Cyg). When eclipse times are available over a long enough time interval, then the apsidal period (e.g., the time it takes for $\omega$ to change by 360°) can be found from the CPOC diagram.

Mathematically, the apsidal motion can be computed as a result of perturbations of a Keplerian orbit that include a General Relativity (GR) contribution, tidal and rotational contributions, and possibly a third-body contribution. The combination of these results in the expression of the rate of change of the argument of periastron,

$$\dot{\omega} = \dot{\omega}_{\mathrm{GR}} + \dot{\omega}_N + \dot{\omega}_3. \quad (1)$$

Analytic expressions exist with which the various $\dot{\omega}$ components can be computed. These expressions may depend on the orbital parameters (e.g., the mean period, the mean eccentricity, and the component masses) and on the stellar parameters (e.g., the stellar radii and the internal density profiles). Depending on the situation, $\dot{\omega}_{\mathrm{GR}}$ may dominate, in which case the binary could be a useful test of GR. In other situations, $\dot{\omega}_{\mathrm{GR}} \ll \dot{\omega}_N$, in which case the rate of apsidal motion can be useful as a probe of the internal structure of the stars.

Generally speaking, in previous studies of apsidal motion, all of the available eclipse times for a given binary have been condensed into a single measurement of $\dot{\omega}$ (e.g., Gimenez & Garcia-Pelayo 1983). For binaries where $\dot{\omega}_N$ dominates, this leads to unavoidable degeneracies because each star contributes to the apsidal motion. Nevertheless, useful information has been found from binaries with roughly equal components, where the hope is that each star has similar interior structures (see the review paper by Torres et al. 2010).

In the ideal case, a model would predict the times of primary and secondary eclipses given the orbital and stellar parameters, which would include a parameterization of the internal density profiles. Then, for a given binary, the observed times of the primary and secondary eclipses (with their uncertainties) can be modeled, and the best-fitting orbital and stellar parameters can be found using standard least-squares techniques. In this paper, we discuss how we use the eclipsing light curve (ELC) program (Orosz & Hauschildt 2000) as such a model. The Newtonian equations of motion with additional force terms to account for GR contributions and tidal distortions are integrated, yielding precise sky positions as a function of time. Times of mid-eclipse and instantaneous orbital elements can then be computed as a function of time.

This paper is organized in the following manner: In Section 2 we discuss the analytic formula used to find $\dot{\omega}$ for the GR, tidal, and third-body cases. In Section 3 we discuss how the ELC code is used to find eclipse times given the orbital and stellar parameters. This model can also be used to find the apsidal period for a given binary, and in Section 5 we validate the ELC model by comparing model-computed values of $\dot{\omega}$ to the standard formula. We conclude with a short summary.

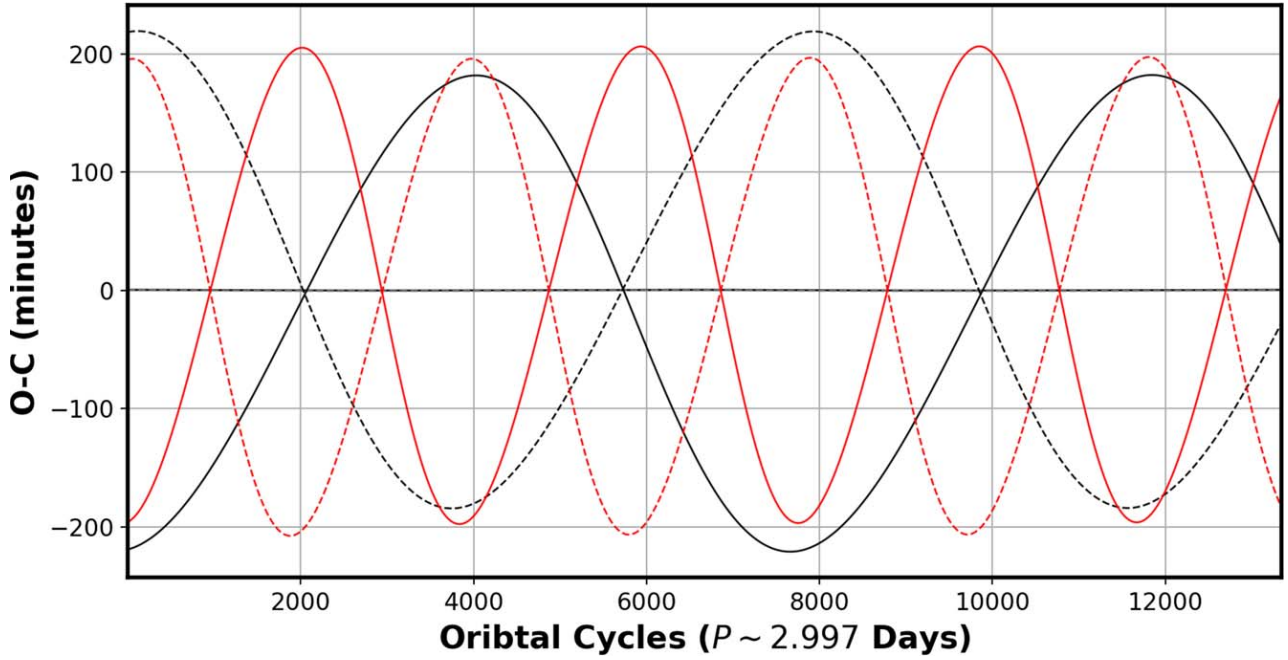

**Figure 1.** CPOC diagram for simulated binary systems resembling Y Cyg, showing apsidal motion due to tides. The solid lines are the times of the primary eclipse, and the dashed lines are the times of the secondary eclipse. The black curves represent the model with the nominal value of the apsidal constants. The red curves are for a system with doubled apsidal constants compared to the black curves, and the frequency of the apsidal precession is therefore doubled. The black line(s) at zero represent a model with apsidal constants of effectively 0, indicating no precession and adherence to a linear ephemeris.

## 2. Analytic Formulae for Contributions to Apsidal Motion

### 2.1. General Relativity Contribution

The rate of apsidal advance in an orbit in GR is given by Equation (10) in Barker & O'Connell (1975a),

$$\dot{\omega} = \frac{3G^{2/3} n^{5/3} (M_1 + M_2)^{2/3}}{c^2 (1 - e^2)}, \quad (2)$$

where $n = 2\pi/P$ is the mean daily motion. Substituting, we obtain

$$\dot{\omega} = \left[\frac{3 G^{2/3}(M_1 + M_2)^{2/3}}{c^2(1-e^2)}\right]\left(\frac{2\pi}{P}\right)^{5/3}, \quad (3)$$

where the units of $\dot{\omega}$ are radians per second. We multiply by the orbital period $P$ to obtain radians per cycle, then by $180/\pi$ to obtain degrees per cycle,

$$\dot{\omega} = \left[\frac{540(2\pi)^{5/3} G^{2/3}}{\pi c^2}\right]\left(\frac{1}{1-e^2}\right)\left(\frac{M_1 + M_2}{P}\right)^{2/3}. \quad (4)$$

In the above equation, the units of the masses are kilograms and the units of the period are seconds. To convert into solar masses and days, respectively, we let $M_\odot$ be the solar mass in kg and $s = 86,400$ be the number of seconds in a day. We then have

$$\dot{\omega} = \left[\frac{540(2\pi)^{5/3}}{\pi c^2}\left(\frac{GM_\odot}{s}\right)^{2/3}\right]\left(\frac{1}{1-e^2}\right)\left(\frac{m_1 + m_2}{P_d}\right)^{2/3} \quad (5)$$

in units of degrees per cycle, where $m_1$ and $m_2$ are the masses of the binary components in solar masses, and $P_d$ is the period in days. The combination $GM_\odot$ is known more accurately than either $G$ or $M_\odot$ are alone,

$$GM_\odot = \frac{k^2 A^3}{s^2}, \quad (6)$$

where $A = 149597870700.0$ is the number of meters in an Astronomical Unit (an exact number by definition), and $k = 0.01720209895$ radians per day is the Gaussian gravitational constant (Clemence 1965). We finally arrive at the well-known formula for $\dot{\omega}_{\rm GR}$ (e.g., Gimenez 1985), where the units are in degrees per cycle,

$$\dot{\omega} = \left[\frac{540(2\pi)^{5/3} k^{2/3} A^2}{\pi c^2 s^2}\right]\left(\frac{1}{1-e^2}\right)\left(\frac{m_1 + m_2}{P_d}\right)^{2/3}$$
$$= (5.447127276 \times 10^{-4})\left(\frac{1}{1-e^2}\right)\left(\frac{m_1 + m_2}{P_d}\right)^{2/3}. \quad (7)$$

The coefficient in this approximation is an exact expression from fundamental constants. From the formula, we see that systems with higher stellar masses and shorter periods will have faster rates of apsidal motion due to GR effects.

### 2.2. Tidal (Newtonian) Contribution

In Newtonian gravity, the orbit of two bound point masses is given by the well-known Kepler equations. The orbit is closed, and the orientation of the semimajor axis (the line of apsides) remains fixed. Real stars are not point masses, and departures from spherical symmetry due to rotation and/or tides give rise to small nonradial forces that cause the orbital elements to change with time (see the Lagrange planetary equations). In most cases, the rate of change of the line of apsides (characterized by the so-called argument of periastron $\omega$) is highest and therefore most readily observable.

From Roy (1978), a form of the Lagrange planetary equations is

$$\frac{da}{dt} = \frac{2}{na}\frac{\partial S}{\partial \chi} \quad (8)$$

$$\frac{de}{dt} = \frac{1}{na^2 e}\left[(1-e)^2)\frac{\partial S}{\partial \chi} - (1-e^2)^{1/2}\frac{\partial S}{\partial \omega}\right] \quad (9)$$

$$\frac{d\chi}{dt} = -\frac{(1-e^2)}{na^2 e}\frac{\partial S}{\partial e} - \frac{2}{na}\frac{\partial S}{\partial a} \quad (10)$$

$$\frac{d\Omega_N}{dt} = \frac{1}{na^2(1-e^2)^{1/2}\sin i}\frac{\partial S}{\partial i} \quad (11)$$

$$\frac{d\omega}{dt} = \frac{(1-e^2)^{1/2}}{na^2 e}\frac{\partial S}{\partial e} - \frac{\cot i}{na^2(1-e^2)^{1/2}}\frac{\partial S}{\partial i} \quad (12)$$

$$\frac{di}{dt} = \frac{1}{na^2(1-e^2)^{1/2}}\left[\cot i\frac{\partial S}{\partial \omega} - \csc i\frac{\partial S}{\partial \Omega_N}\right], \quad (13)$$

where $n^2 a^3 = G(m_1 + m_2)$ and $\chi = -nT$, with $n = 2\pi/P$ being the mean daily motion. The eccentricity, nodal angle, and inclination are defined by $e$, $\Omega_N$, and $i$, respectively. These expressions and the functional form of $S$ therein determine which of the orbital elements will change over time, and how quickly they change.

Expressions for $\dot{\omega}_N$ due to tidal distortions were first derived by Cowling (1938) and Sterne (1939). Following the revised formulation by Kopal (1978), the expression for the apsidal period $U$ from tides is given by

$$\frac{P}{U} = \frac{\dot{\omega}_N}{2\pi} = c_1 k_{21} + c_2 k_{22}, \quad (14)$$

where the $k_{2i}$ factors are the second-order (quadrupolar) internal structure constants. These internal structure constants are related to the density distribution within the star, and they can be computed from stellar evolution models.

The weighting coefficients $c_i$ are functions of the mass ratio, eccentricity, and relative radii with respect to the orbital separation, and they are given by

$$c_i = r_i^5\left(\frac{m_{3-i}}{m_i}[15g(e) + \gamma_i^2 f(e)] + \gamma_i^2 f(e)\right), \quad (15)$$

where $r_i$ is the fractional radius scaled to the orbital separation ($R_i/a$), the parameters $\gamma_i$ are the ratio of the angular velocity of the axial rotation to that of orbital motion, and $g(e)$ and $f(e)$ are functions of the eccentricity, originally estimated as a power series in $e$ by Sterne (1939), and provided by Bulut et al. (2017) as

$$f(e) = \frac{1}{(1-e^2)^2} \quad (16)$$

$$g(e) = \frac{(8 + 12e^2 + e^4)f(e)^{2.5}}{8}. \quad (17)$$

When we adopt pseudo-synchronous rotation in eccentric orbits, as done in Hut (1981), the maximum angular velocity at periastron can be well approximated as

$$\gamma_i^2 = \frac{(1+e)}{(1-e)^3} = \frac{\omega_P^2}{\Omega_K^2}. \quad (18)$$

A similar recent approach by Bulut et al. (2017) determines the $c_i$ coefficients through a combination of the contributions of rotational distortion and tidal effects,

$$c_i = r_i^5\left[\left(\frac{\Omega_{r,i}}{\Omega_K}\right)^2\left(1 + \frac{m_{3-i}}{m_i}\right)f(e) + \frac{15m_{3-i}}{m_i}g(e)\right]. \quad (19)$$

In the expression for $c_i$ (Equation (19)), $r_i$ is the fractional radius, $m_i$ is the mass, $\Omega_r$ is the angular velocity of the axial rotation for each component $i$, $\Omega_K$ is the Keplerian angular velocity, and $e$ is the orbital eccentricity. Thus, when the Keplerian parameters are known, then the stellar parameters including the rotation rates and the apsidal period can be calculated.

The mean value of the internal structure constants can be derived from the observed value of $\dot{\omega}$ using the expression

$$\bar{k}_{2,\text{obs}} = \frac{1}{c_{21} + c_{22}}\frac{P}{U} = \frac{1}{c_{21} + c_{22}}\frac{\dot{\omega}}{2\pi}, \quad (20)$$

where the $c_{2i}$ coefficients are the same functions of the eccentricity, mass, radius, and separation as in Equation (19).

It is known that the observed mean value of the internal structure constants contains both contributions from the Newtonian and the relativistic effects of apsidal motion. When the constants are combined through the equation

$$\bar{k}_{2,\text{theo}} = \frac{c_{21}k_{21,\text{theo}} + c_{22}k_{22,\text{theo}}}{c_{21} + c_{22}}, \quad (21)$$

the weighted average coefficient $\bar{k}_{2,\text{theo}}$ can be determined. This weighted average is directly comparable with the observed value. However, this historical method fails to constrain the individual constants $k_{21}$ and $k_{22}$ as only the average value is returned, and this method does not work well with binaries with mass ratios $q \napprox 1$.

### 2.2.1. Stellar Spin Axes

We can break down the Newtonian component of apsidal motion into the contributions from tides and rotation:

$$\dot{\omega}_{\text{total}} = \dot{\omega}_{\text{GR}} + \dot{\omega}_{\text{tidal},1} + \dot{\omega}_{\text{tidal},2} \\ + \dot{\omega}_{\text{rot},1}\phi_1 + \dot{\omega}_{\text{rot},2}\phi_2 + \dot{\omega}_{\text{LTT}}.$$

If the spin axes are misaligned, there is an additional contribution to the tidal term in the expression for apsidal motion. The pointing direction of the angular momentum vector of the stars will affect the apsidal motion of the system. This is parameterized in two values, where $\angle \alpha_i$ is the deflection in the plane of the orbit, and $\angle \beta_i$ is the deflection in the plane of the sky. The combination of the two of these can result in any pointing direction for either the primary or secondary star.

A summary of the theory behind misaligned spin axes and the subsequent affect on the secular motion of the apse is given in Shakura (1985). In particular, his Equation (4) gives the rate of change of $\omega$,

$$\frac{d\omega}{dt} = \left(\frac{d\omega}{dt}\right)_E + \bar{\omega} 15 g(e) \left[k_1 r_1^5 \frac{m_2}{m_1} + k_2 r_2^5 \frac{m_1}{m_2}\right]$$
$$- \frac{\bar{\omega}}{\sin^2 i} f(e) \left\{k_1 r_1^5 \left(\frac{\omega_1}{\bar{\omega}}\right)^2 \left(1 + \frac{m_2}{m_1}\right)\right.$$
$$\times \left[\cos\alpha_1(\cos\alpha_1 - \cos\beta_1 \cos i) + \frac{1}{2}\sin^2 i(1 - 5\cos^2\alpha_1)\right]$$
$$+ k_2 r_2^5 \left(\frac{\omega_2}{\bar{\omega}}\right)^2 \left(1 + \frac{m_1}{m_2}\right)$$
$$\left.\times \left[\cos\alpha_2(\cos\alpha_2 - \cos\beta_2 \cos i) + \frac{1}{2}\sin^2 i(1 - 5\cos^2\alpha_2)\right]\right\},$$
(22)

where the $(d\omega/dt)_E$ is the GR contribution. If the spin axes of the stars are aligned, then $\alpha_1 = \alpha_2 = 0$ and $\beta_1 = \beta_2 = i$, where $i$ is the inclination of the orbital plane of the binary.

### 2.3. Third-body Contribution

Although we focus our attention in this paper on binaries with apsidal motion due to tidal or GR effects, for completeness, we mention the possible contribution from a third body, which can lead to observable changes in the CPOC diagram. The addition caused by the third body has to be accounted for before the apsidal motion can be analyzed.

Because the binary orbits the barycenter of the triple system, the observed times of the primary and secondary eclipses can either be early or late because the distance between the observer and the binary changes periodically. The signal in the $O - C$ diagram then superficially resembles a radial velocity curve. Irwin (1952) gives a formula to model this light travel time effect (LTTE) signal. In the massive binary DR Vul, the CPOC signals have been modeled by a combination of an LTTE orbit with an orbital period of $\approx 63$ yr and apsidal motion of the inner binary with an apsidal period $\approx 36$ yr (Wolf et al. 2019; Dimoff 2021).

If the third body is sufficiently close to the binary, the gravitational perturbations can lead to changes in the orbital period as well as in the precession of the orbit, and these changes can even dwarf the LTTE. Several binaries with large eclipse-timing variations (ETVs) due to a third body have been discovered using data from the Kepler and Transiting Exoplanet Survey Satellite (TESS) missions. Generally speaking, the ETV signal seen in the $O - C$ diagram from dynamical interactions is complex. Borkovits et al. (2011) and Baycroft et al. (2023; their Appendix A) give expressions for $\dot{\omega}$ including effects from a third body. Although these dynamical interactions can produce measurable effects in the $O - C$ on both short and long timescales, the short-period low-amplitude variations are less important for apsidal motion studies. The long-period perturbations in the apsidal motion, however, can substantially alter the tidal and relativistic effects (see Naoz 2016, or Borkovits et al. 2020 for a recent example).

### 3. Modeling Eclipse Times Using ELC

We now discuss our forward model, which can produce the times of primary and secondary eclipses given the stellar parameters and initial orbital parameters. It is based on the ELC code (Orosz & Hauschildt 2000; Orosz et al. 2019). The code is general, and the light and velocity curves of a variety of binary and three-body systems can be modeled directly. In the mode that we describe here, the observed times of eclipse can be fitted, which is useful in situations without access to the light curves.

The basic outline of a photodynamical code like ELC is relatively straightforward. Given the masses of two (or more) bodies and initial conditions (positions and velocities) for these bodies, the equations of motion are integrated, yielding the sky positions as a function of time. From the time series of the positions, it is easy to find the times when any two bodies are at conjunction. It can easily be checked if an eclipse occurs at or near conjunction when given the radius of each body.

For convenience, ELC has a Keplerian-to-Cartesian converter (based on the algorithms given in Murray & Dermott 1999), where six orbital elements (the period $P$, the time of periastron passage $T$, the eccentricity $e$, the argument of periastron $\omega$, the inclination $i$, and the nodal angle $\Omega$) uniquely determine six phase-space coordinates for each star (e.g., $x$, $y$, $z$, $v_x$, $v_y$, and $v_z$). The $x$, $y$ plane is the sky plane, and the $z$-axis points toward the observer. The inverse transformation (the Cartesian-to-Keplerian converter) is available, where six phase-space coordinates give a unique set of orbital elements.

The equations of motion are the usual force equations for the two-body problems, plus additional force terms that arise from tidal distortions and force terms from the full GR treatment (Mardling & Lin 2002),

$$\ddot{\bm{r}} = -\frac{G(m_1 + m_2)}{r^3}\bm{r} + \bm{f}_{\text{QD},1} + \bm{f}_{\text{QD},2} + \bm{f}_{\text{rel}} + \bm{f}_3. \quad (23)$$

The acceleration due to the quadrupole moment of body 1 is a combination of its spin distortion and tidal distortion produced by the presence of the companion,

$$\bm{f}_{\text{QD},1} = \frac{R_1^5(1 + m_2/m_1)k_{21}}{r^4}$$
$$\times \left(\left[5(\bm{\Omega}_1 \cdot \hat{\bm{r}})^2 - \Omega_1^2 - \frac{6Gm_2}{r^3}\right]\hat{\bm{r}}\right.$$
$$\left. - 2(\bm{\Omega}_1 \cdot \hat{\bm{r}})\bm{\Omega}_1\right), \quad (24)$$

where $k_{21}$ is the apsidal motion constant of body 1, $\bm{\Omega}_1$ is the ratio of the rate of stellar rotation and the orbital rotation, and $\hat{\bm{r}}$ is a unit vector in the direction of $r$. A similar expression exists for the tidal distortion of body 2 due to body 1.

The orbital acceleration due to the relativistic potential of the binary (Kidder 1995) is given by

$$\bm{f}_{\text{rel}} = -\frac{Gm_{12}}{r^2 c^2}$$
$$\times \left(\left[(1 + 3\eta)\dot{\bm{r}} \cdot \dot{\bm{r}} - 2(2 + \eta)\frac{Gm_{12}}{r} - \frac{3}{2}\eta\dot{r}^2\right]\hat{\bm{r}}\right.$$
$$\left. - 2(2 - \eta)\dot{r}\dot{\bm{r}}\right), \quad (25)$$

where $m_{12} = m_1 + m_2$, $c$ is the speed of light, and $\eta = m_1 m_2/(m_{12})^2$. The force due to a third body (where applicable) is given by

$$\bm{f}_3 = Gm_3(\bm{\beta}_{23} - \bm{\beta}_{13}), \quad (26)$$

where $\bm{\beta}_{ij} = \bm{r}_{ij}/r_{ij}^3$ is a ratio of the relative positions of the bodies in the system.

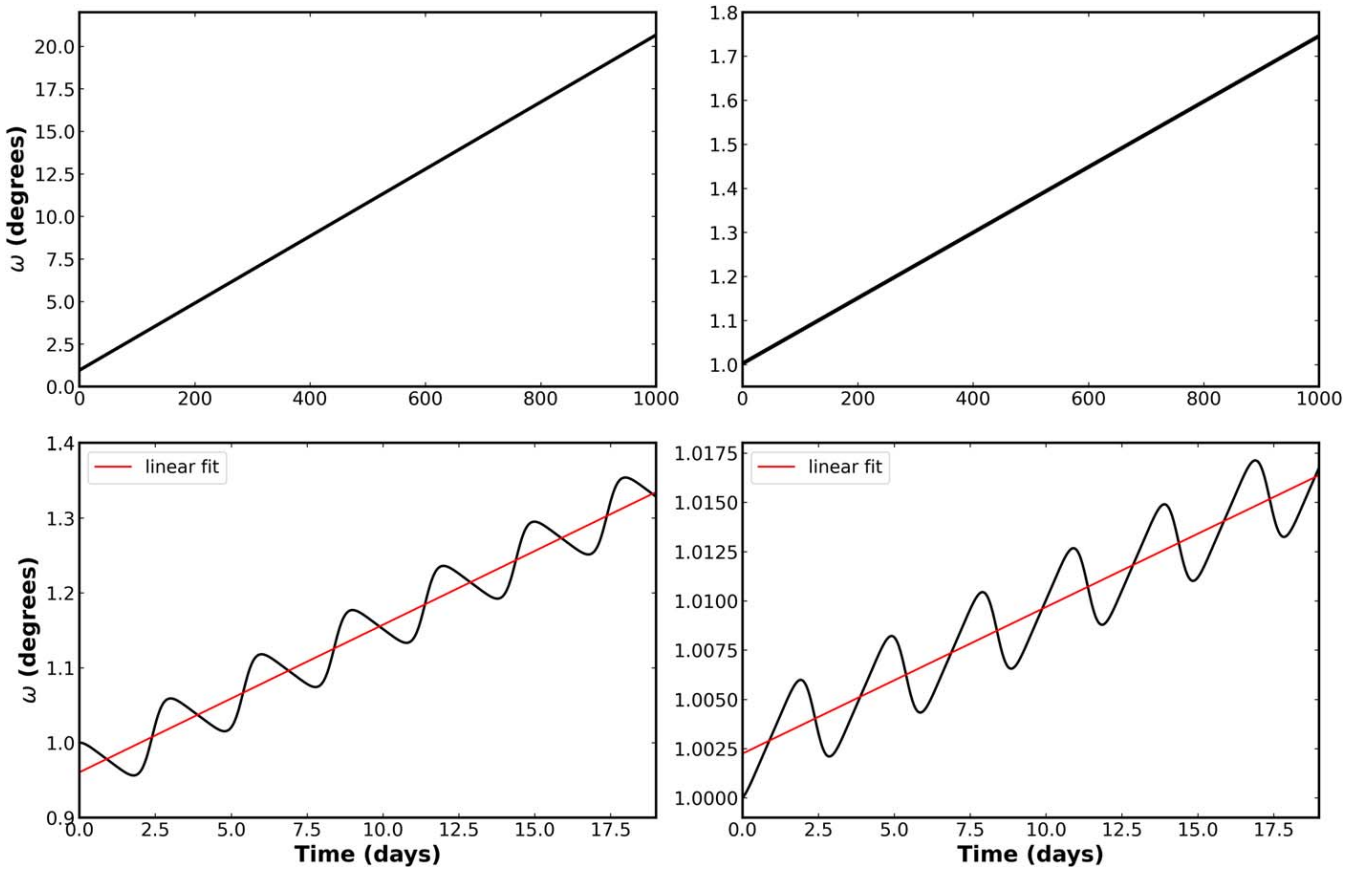

**Figure 2.** Advance of the argument of periastron over time for a Y Cyg-like system. The top panels show the linear progression of $\omega$ with time considering tides and GR independently (left and right, respectively). In the bottom panels, the red line is the linear regression. The observed oscillations from the linear trend in the bottom panels occur on an orbital timescale (for Y Cyg ∼ 3 days). The shapes of the curves are sinusoidal to first order, and the amplitudes are related to the length of the apsidal cycle.

To find the positions of all the bodies at any given time, the equations of motion are integrated using a symplectic twelfth-order Gaussian Runge–Kutta (GRK) integration routine based on the methods from Hairer et al. (2006). In our implementation, one can solve only the purely Newtonian equations (the forces are only from point masses), the Newtonian equations plus the tidal forces, the Newtonian equations plus the GR forces, or the equations with all three contributions. As noted previously, if the additional force terms due to tides or due to GR effects are included, the orbit will not be closed, and the orbital elements will change with time. The Cartesian-to-Keplerian converter can be used to compute time series of the orbital elements. Of interest here is the time series for the argument of periastron $\omega(t)$.

## 4. Comparing Derived Apsidal Rates with Analytic Formula

The analytic formula discussed in Section 2 gives the apsidal rate, namely $\dot\omega$. For a model that can produce a time series for $\omega(t)$, the rate of change of $\omega$ (e.g., $\dot\omega$) should be easy to compute and can be compared to the analytic formula. However, in this case, there are some subtle issues. Figure 2 displays the argument of periastron ($\omega$) over time for a system resembling Y Cyg considering tides and GR independently. For this calculation, $\omega_0 = 1°$. The system is precessing, and over the course of 1000 days, $\omega$ changed by about 20° due to tidal effects. However, when the graph is examined more closely, small small-scale oscillations become visible with a period equal to the binary orbital period. The amplitude of these oscillations depends on the apsidal period, and in the limit where there is no apsidal motion (e.g., when $\dot\omega = 0$), the oscillations must vanish. At each time step, six phase-space coordinates give a unique set of orbital parameters (including $\omega$).

However, because the underlying motion is not Keplerian, higher-order oscillations are noticeable. The equations of motion with the corrections provide an orbit-averaged force. The effect of that force depends on the instantaneous separation of the bodies, so the actual speed of the star changes in a slightly different manner relative to the average Keplerian. Given these oscillations, the value of $\dot\omega$ from a given model is computed by using a linear fit to the $\omega(t)$ curve. More reliable values of omega dot are obtained from averaging over a longer time-span that contains many orbital periods.

From a collection of eclipsing binaries exhibiting rapid apsidal motion (Claret & Willems 2002), we select 15 prototypical binary systems to use as a test set for our model. We simulate the orbital motion of these systems each for a set number of days relative to an apsidal cycle, taking into account their physical parameters including the mass, radius, apsidal constants, and orbital spin-axis orientations of both components. A summary of the input parameters for each system is presented in Table 1 (for assumed rotationally aligned systems). We do not claim that these are the actual physical parameters of the selected systems, but they are similar enough to their physical values to represent realistic binaries.

For each of the 15 binaries, we ran 200,000 forward models for three situations: (i) GR perturbations only, (ii) tidal

**Table 1**
Adopted Input Orbital and Physical Parameters for the Selected Binaries Exhibiting Apsidal Motion

| System | $P$ (days) | $e$ | $i$ | $m_1, m_2$ ($M_\odot$) | $r_1, r_2$ ($R_\odot$) | $\Omega_1, \Omega_2$ | $rk_{21}, rk_{22}$ | References |
|---|---|---|---|---|---|---|---|---|
| AG Per | 2.0288769 | 0.0709 | 81.40 | 5.35 | 2.995 | 1.156 | 0.0072 | 0,1 |
|   |   |   |   | 4.489 | 1.1497 | 1.156 | 0.0072 |   |
| DI Her | 10.550170 | 0.4895 | 89.30 | 5.17 | 2.681 | 1.000 | 0.0077 | 27,28,29,30,32 |
|   |   |   |   | 4.524 | 2.478 | 1.000 | 0.0077 |   |
| EM Car | 3.4149023 | 0.0119 | 81.50 | 21.394 | 8.371 | 0.999 | 0.0080 | 0,2,22,35 |
|   |   |   |   | 22.883 | 9.347 | 1.048 | 0.0029 |   |
| IQ Per | 1.7436292 | 0.0763 | 89.30 | 3.504 | 2.445 | 1.169 | 0.0044 | 0,3,4,5,6 |
|   |   |   |   | 1.730 | 1.499 | 1.169 | 0.0044 |   |
| QX Car | 4.4781279 | 0.279 | 85.70 | 6.240 | 4.291 | 1.000 | 0.0071 | 0,7,8 |
|   |   |   |   | 8.460 | 4.053 | 1.000 | 0.0071 |   |
| V1647 Sgr | 3.2828505 | 0.4142 | 90.00 | 2.184 | 1.832 | 1.000 | 0.0042 | 0,9,10,11 |
|   |   |   |   | 1.967 | 1.667 | 1.000 | 0.0042 |   |
| V526 Sgr | 1.9194849 | 0.2199 | 89.10 | 2.206 | 1.880 | 1.712 | 0.0038 | 22,23,24,35 |
|   |   |   |   | 1.680 | 1.820 | 1.269 | 0.0020 |   |
| Y Cyg | 2.9968537 | 0.1462 | 86.47 | 17.790 | 5.525 | 1.358 | 0.0095 | 0,12,13,14,35 |
|   |   |   |   | 18.296 | 5.784 | 1.358 | 0.0157 |   |
| CW Cep | 2.7294954 | 0.0292 | 81.80 | 12.932 | 5.521 | 1.029 | 0.0053 | 0,15,16,17,35 |
|   |   |   |   | 11.898 | 5.095 | 1.019 | 0.0125 |   |
| DR Vul | 2.2512153 | 0.0945 | 88.30 | 13.203 | 4.801 | 0.796 | 0.0063 | 25,26,35 |
|   |   |   |   | 12.189 | 4.336 | 0.992 | 0.0150 |   |
| GG Lup | 1.8496919 | 0.1546 | 78.00 | 4.106 | 2.380 | 1.382 | 0.0070 | 0,18 |
|   |   |   |   | 2.504 | 1.726 | 1.382 | 0.0070 |   |
| U Oph | 1.6773459 | 0.00 | 87.86 | 5.090 | 3.440 | 1.000 | 0.0053 | 0,5,17,19 |
|   |   |   |   | 4.580 | 3.050 | 1.000 | 0.0053 |   |
| V478 Cyg | 2.881 | 0.0158 | 78.00 | 16.60 | 7.430 | 1.032 | 0.0056 | 29,31,33,34 |
|   |   |   |   | 16.30 | 7.430 | 1.032 | 0.0056 |   |
| V760 Sco | 1.6697772 | 0.0113 | 85.00 | 3.921 | 2.852 | 1.023 | 0.0044 | 0,20,11 |
|   |   |   |   | 2.545 | 1.854 | 1.023 | 0.0044 |   |
| $\zeta$ Phe | 1.66977 | 0.0116 | 89.30 | 3.908 | 2.835 | 3.346 | 0.0077 | 0,21,15,7 |
|   |   |   |   | 2.536 | 1.885 | 3.436 | 0.0077 |   |

**Note.** Input data were collected from (0) Claret & Willems (2002), (1) Gimenez & Clausen (1994), (2) Andersen & Clausen (1989), (3) Burns et al. (1996), 4) Caton & Burns (1993), (5) Andersen (1991), (6) Lacy & Frueh (1985), (7) Andersen et al. (1983), (8) Gimenez et al. (1986), (9) Clausen et al. (1977), (10) Andersen & Gimenez (1985), (11) Wolf (2000), (12) Hill & Holmgren (1995), (13) Simon et al. (1994), (14) Holmgren et al. (1995), (15) Claret & Gimenez (1993), (16) Claret & Gimenez (1991), (17) Popper & Hill (1991), (18) Andersen et al. (1993), (19) Kaemper (1986), (20) Andersen et al. (1985), (21) Clausen (1996), (22) Wolf & Zejda (2005), (23) Lacy (1997), (24) Lacy (1993), (25) Bozkurt & Değirmenci (2007), (26) Wolf et al. (2019), (27) Albrecht et al. (2009), (28) Popper (1982), (29) Marcussen & Albrecht (2022), 30) Anderson & Winn (2022), (31) Claret et al. (2021), (32) Claret et al. (2010), (33) Claret & Giménez (2010), (34) Pavlovski et al. (2018), and (35) Dimoff (2021).

perturbations with aligned spin axes, and (iii) tidal perturbations with misaligned spin axes. The initial value of $\omega$ was always $1°$, and the values of the other orbital elements and the stellar masses and radii were drawn from normal distributions that represent typical measurement uncertainties on these quantities. The model values of $\dot{\omega}$ for each binary system are determined from a linear regression of $\omega(t)$, and are compiled in Table 2. We compare these values to results from the respective analytic formula as a fractional percent difference, and discuss each of these three cases in turn.

### 4.1. Accuracy of the General Relativity Contribution

A histogram of the error distributions for the GR case for each of our selected 15 systems is shown in Figure 3, limited to the range $-0.002\%$ to $0.002\%$. While the combined distributions are not centered exactly on zero, each individual distribution is approximately symmetric. Systems with lower eccentricity (e.g., CW Cep, EM Car, V478 Cyg, or U Oph) exhibit wider histograms, as well as peaks or horns close to the ends of the distributions. This is the result of the near-ambiguity in the value of the argument of periastron $\omega$ in a nearly circular orbit.

When only GR is taken into account, the model apsidal rates agree quite well with the formula given by Equation (7). The percent errors are $\sim 5 \times 10^{-4}$ on average. We note at this level, there are not enough digits in the coefficients given in Equation (7) from Gimenez (1985) for a meaningful comparison with our models.

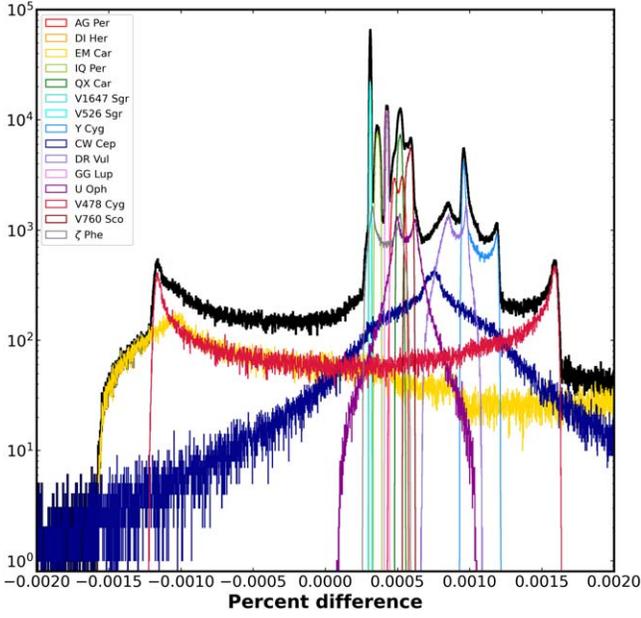

**Figure 3.** Composite histogram of models representing the error distributions when only GR is considered. The black curve is the sum of all distributions.

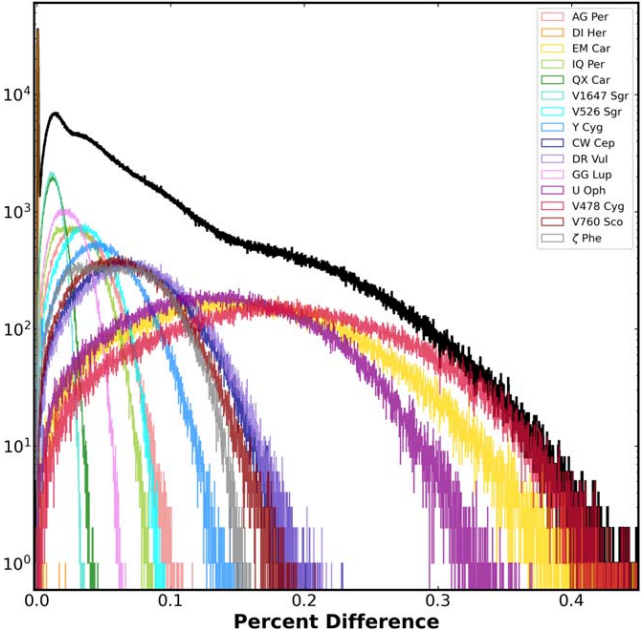

**Figure 4.** Composite histogram of models representing the error distributions when only tides are considered. The black curve is the sum of all distributions.

### 4.2. Accuracy of the Tidal Contribution

Histograms of the error distributions for each of our selected 15 systems considering the tidal contribution are shown in Figure 4 for the aligned case and in Figure 5 for the misaligned case. The analytic approach to modeling tides taken by Sterne (1939; e.g., Equation (14)) works well. In this case, we find that the relative errors are acceptably small, within 0.4%. For the misaligned case (Equation (22)), the percent errors are larger, ~2%.

### 5. Discussion

Our implementation of GR is very accurate when compared to the analytic formula (Equation (7)). The differences are small but systematically positive, where the median percent difference is ≈0.00067. For comparison, Weinberg (1972) gives the rate of precession as

$$\Delta\varphi = \left(\frac{6\pi MG}{L}\right) \text{(radians revolution}^{-1}\text{)}, \quad (27)$$

where $L = (1 - e^2)a$ is a measure of the eccentricity of the orbit, the so-called semilatus rectum. This formula is exact in the Schwarzschild metric, and agrees with Barker & O'Connell (1975b) when appropriate substitutions are made. The agreement between Equation (5) and our model indicates that the force equations account quite well for the GR perturbations.

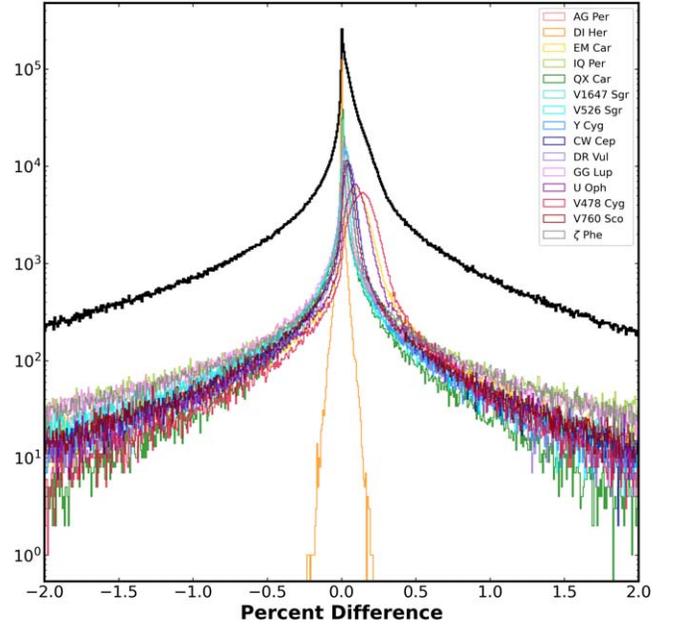

**Figure 5.** Composite histogram of models representing the error distributions when only tides are considered. In this case, the primary star is misaligned by a set of angles: $axis_i = +40°$ through $+140°$, and $axis_b = -50°$ through $+50°$. The black curve is the sum of all distributions.

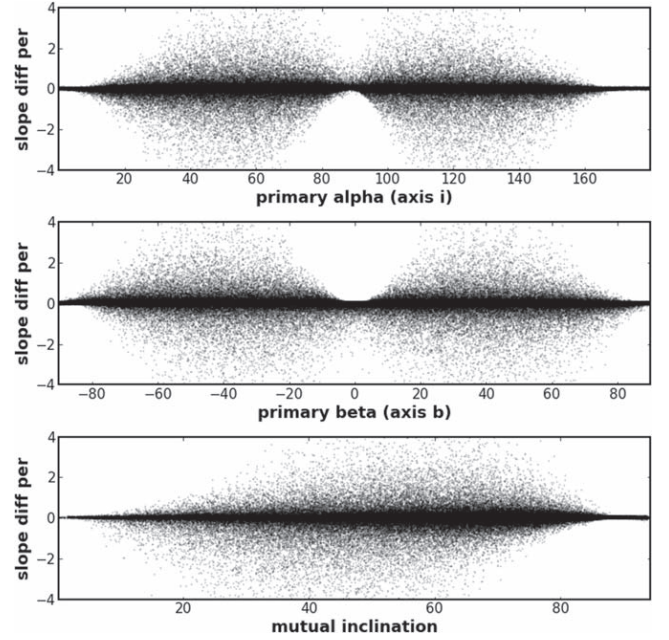

**Figure 6.** Percent differences between the computed slope and the theoretical value for a Y Cyg-like system for a set of 200,000 simulations for a range of primary stellar orbital spin axes and resulting mutual inclinations. The slanted trend in the mutual inclination plot may arise because the inclination of the orbital system is not being perfectly edge-on, i.e., $i \neq 90°$.

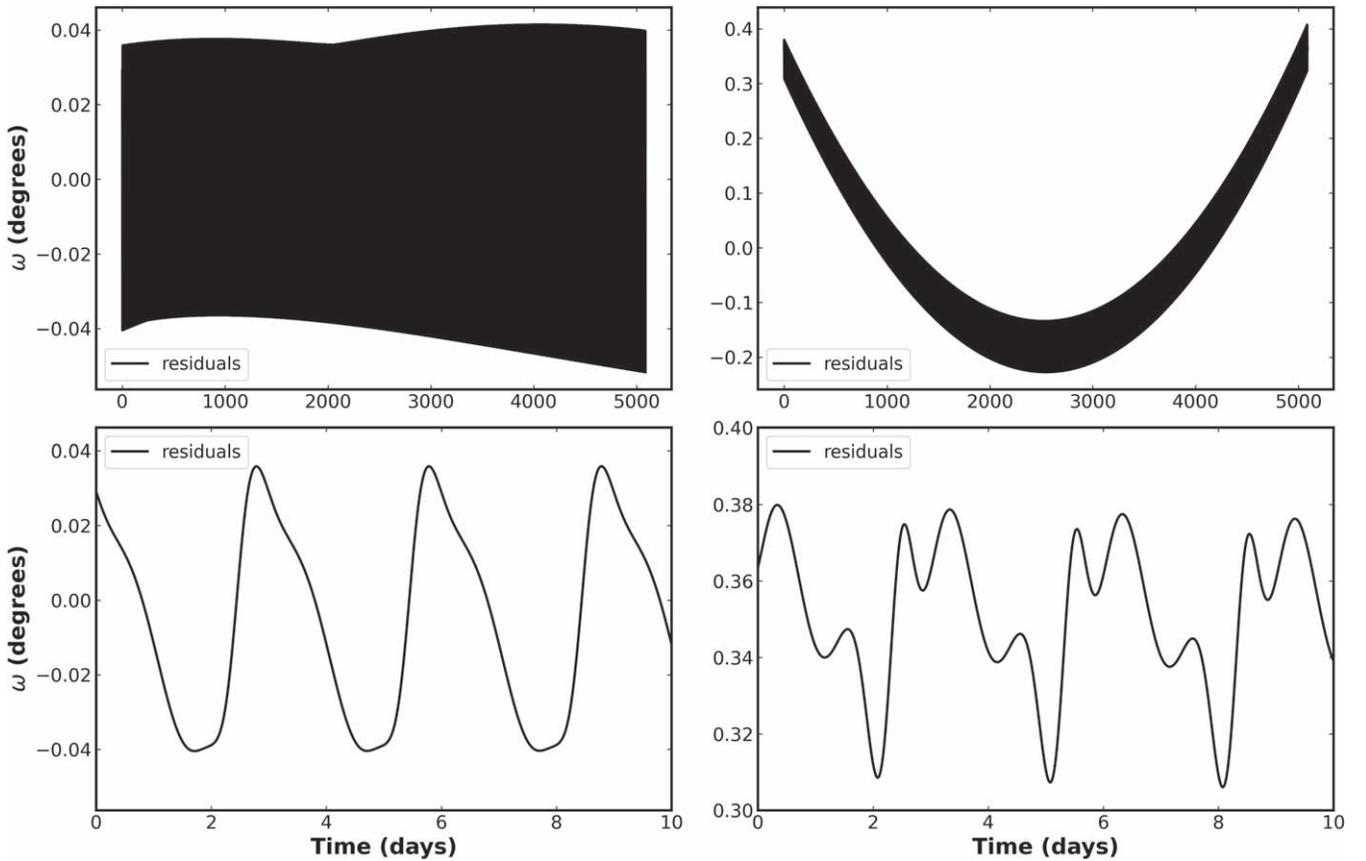

**Figure 7.** Residuals from a linear fit to the advance of the argument of periastron in two cases of axial misalignment for a Y Cyg-like system. The left panels represents a system close to alignment; the axial deflections result in a slightly different pattern of the oscillations, but maintain a linear trend. The right panels represent a system far from axial alignment; the oscillations have more structure, and the overall trend is nonlinear.

While not as good as GR, our modeled apsidal rates for the tidal contribution with an aligned stellar spin axes are accurate compared to the analytic formula (Equation (14)). Again, the differences are systematically positive, with a median percent difference value of $\approx 0.013$. Unlike the GR case, there is an approximation in the formula for $\dot{\omega}$. Sterne (1939) uses the $n = 2$ (second-order) term in the tidal potential, which is proportional to $(r^2/a^3)$. The additional force terms in our model are presumably derived from a treatment that uses the $n = 2$ term in the disturbing potential. Apparently, in this case, the solutions to the differential equations do not reproduce the analytic results as faithfully as they do in the GR case. According to Sterne (1939), when the third-order term is included in the tidal potential, $\dot{\omega}$ scales with $(r/a)^7$. For the binaries considered here, $(r/a)^7$ is only a few percent of $(r/a)^5$ in Equation (19).

When the axial misalignment of the stars is taken into account, the accuracy is not as good, although the histograms of the percent differences are closer to symmetric about zero. For each star, we establish an axis deflection $\alpha_i$ as the spin–orbit inclination, and axis $\beta_i$ as the inclination with respect to the plane of the sky (Equation (22)). These axial inclination parameters for each star can define any spin-axis orientation for a binary. For a system where the spin axes of the stars are aligned with the angular momentum axis of the orbit to within about 15°, Equation (22) is accurate to within a few percent.

As seen in Figure 6, the largest relative errors occur at mutual inclinations of $\sim 45°$, representing the maximum allowed asymmetry as the tidal bulge of the companion changes hemispheres. Furthermore, the relative errors are minimized at mutual inclinations of 0° and 90°, where in the former case, the spin axis is perpendicular to the plane of the orbit, and in the latter case, the spin axis is within the plane of the orbit. One simulated system each of these regimes is plotted in Figure 7.

We fit a linear regression to the line of $\omega$ versus $t$ and compute the residuals. In systems close to the aligned case (mutual inclination $\approx 0$), a linear trend is recovered, visible in the top left panel of Figure 7. For extreme cases, the $\omega(t)$ is highly nonlinear. In these extreme cases, where the mutual inclination $\napprox 0$ (where the computed relative error compared to the analytic formula is large), we find a second-order trend with the $\omega(t)$ curve. This can be seen in the top right panel of Figure 7.

Mardling & Lin (2002) point out that if the spin axis of one of the stars is not aligned, there will be a torque on that star. If there is a torque on that star, then the direction of the spin axis will change ($\dot{\mathbf{\Omega}}_r \propto \mathbf{r} \times \mathbf{f}_{QD}$). The expression for the force given in Equation (24) contains $\Omega_r$, so that in principle, if that vector changes with time, it could add complications to the system. However, our numerical experiments show that $\dot{\mathbf{\Omega}}_r$ is very small in most practical cases. Furthermore, Shakura (1985) shows that the inclination $i$ and nodal angle $\Omega_N$ also change

**Table 2**
Representative Apsidal Rates of Advance and Apsidal Periods as a Result of Their Physical and Orbital Parameters Used in Our Modeling

| System | $\dot{\omega}_{GR}$ (deg cycle$^{-1}$) | $U_{GR}$ (yr) | $\dot{\omega}_{Tidal}$ (deg cycle$^{-1}$) | $U_{Tidal}$ (yr) |
|---|---|---|---|---|
| AG Per | 0.00161 | 1240.8 | 0.03714 | 53.8 |
| DI Her | 0.00068 | 15358.6 | 0.00069 | 14999.0 |
| EM Car | 0.00300 | 1120.9 | 0.17029 | 19.8 |
| IQ Per | 0.00113 | 1515.2 | 0.03220 | 53.4 |
| QX Car | 0.00147 | 2987.2 | 0.01346 | 327.8 |
| V1647 Sgr | 0.00076 | 4233.2 | 0.01168 | 276.9 |
| V526 Sgr | 0.00091 | 2073.1 | 0.01277 | 148.3 |
| Y Cyg | 0.00293 | 1008.1 | 0.05303 | 55.7 |
| CW Cep | 0.00238 | 1128.8 | 0.07803 | 34.5 |
| DR Vul | 0.00277 | 800.0 | 0.08160 | 27.2 |
| GG Lup | 0.00130 | 1397.5 | 0.02499 | 72.9 |
| U Oph | 0.00179 | 926.4 | 0.15210 | 10.9 |
| V478 Cyg | 0.00276 | 1027.6 | 0.20082 | 14.1 |
| V760 Sco | 0.00170 | 1000.6 | 0.07117 | 23.9 |
| ζ Phe | 0.00134 | 1223.9 | 0.06609 | 24.9 |

**Note.** For all systems except for DI Her, the rate of advance due to GR is much slower than the tidal rate.

with time, and gives corresponding formulae for $di/dt$ and $d\Omega_N/dt$. Given the changes in the other orbital elements, it is not surprising that the $\omega$ versus $t$ curve becomes nonlinear on a shorter timescale. Indeed, Equation (22) for $d\omega/dt$ includes a term for the inclination. If $di/dt$ is nonzero, there are additional contributions to $d\omega/dt$.

## 6. Conclusion

The precession of the line of apsides in a binary orbit is driven by perturbations from GR, tidal, and third-body interactions. These perturbations depend on the physical and orbital properties of the systems, including the orbital period, eccentricity, stellar mass, internal structure constants, and relative rotation speeds of each star, and it is affected by the angular deflection of the stellar rotation axes. We investigate the effectiveness of the ELC software in modeling the apsidal motion of 15 realistic binary systems. Modifying Newton's equations of orbital motion to include dynamical perturbations from tides, GR, axial misalignment, and possible third-body effects, we run a set of forward models and compute the rates of apsidal motion, and compare them to the corresponding analytical formulae. Our results indicate an extremely good agreement between for the GR contribution, and a good agreement between our modeling and theoretical predictions for the tidal contribution to the apsidal motion. Despite this agreement, inconsistencies remain in the case considering tides with misaligned rotation axes. These functional degeneracies between derived orbital and physical parameters will be addressed in a follow-up study.

This approach to numerically modeling the tidal and GR forces is a fast and precise way to model apsidal motion in binary (or higher-order multiple) stellar systems. We plan to apply this technique to real eclipse-timing data and compare our results in a future work, and we aim to compute refined orbital parameters, internal structure constants, and stellar rotation axis alignments.

## Acknowledgments

This project has received funding from the European Union's Horizon 2020 research and innovation program under grant agreement No. 101008324 (ChETEC-INFRA). Support in funding comes from the State of Hessen within the Research Cluster ELEMENTS (Project ID 500/10.006). This work made use of the EMMY supercomputer, provided by the North German Supercomputing Alliance (Norddeutscher Verbund für Hoch- und Höchstleistungsrechnen—HLRN), hosted at Georg-August-Universität Göttingen.

*Software:* ELC (Orosz & Hauschildt 2000), (Orosz et al. 2019), Numpy (Harris et al. 2020), Matplotlib (Hunter 2007).

## ORCID iDs

Alexander J. Dimoff  https://orcid.org/0009-0007-3458-0401
Jerome A. Orosz  https://orcid.org/0000-0001-9647-2886